\documentstyle[pre,twocolumn,aps,epsf,floats]{revtex} 

\begin{document} 

\draft

\title{Large scale simulations of the 
Zhang sandpile model}

\author{S. L\"ubeck\cite{SvenEmail} }
\address{Theoretische Tieftemperaturphysik, 
Gerhard-Mercator-Universit\"at Duisburg,\\ 
Lotharstra{\ss}e 1, 47048 Duisburg, Germany \\}
\author{\vskip -2\baselineskip\small(Received 26 February 1997)\break}
\author{\parbox{14cm}{\small
        \quad We consider the non Abelian sandpile model
	introduced by Y.-C.~Zhang on a two-dimensional square lattice.
	The static and dynamical properties of the model are investigated
	and compared to the Abelian sandpile model of Bak, Tang
	and Wiesenfeld.
	A detailed analysis which takes the finite size effects
	into account yields that the exponents of the avalanche probability
	distribution are the same as in the Abelian model. 
\hfill\break
\leftline{PACS number: 05.40.+j}}}
\address{\vskip +1.5\baselineskip}

\maketitle

\setcounter{page}{1}
\markright{\rm
accepted for publication in  Phys. Rev.~E {\bf 56}, ? (1997)
  }
\thispagestyle{myheadings}
\pagestyle{myheadings}

\section{Introduction}

The idea that an externally driven physical system with 
many degrees of freedom can be self-organized critical (SOC) was 
introduced few years ago by Bak, Tang and Wiesenfeld (BTW) 
and realized theoretically using a stochastic cellular 
automaton \cite{BAK}.
The original BTW model belongs to the Abelian sandpile models
\cite{DHAR_2}. 
Here the sequence of relaxation processes is described by operators
which satisfy a commutative algebra.
This property allows the analytical calculation of some 
features of the system in the steady 
state \cite{DHAR_2,MAJUM_1,PRIEZ_1,IVASH_1}.
A continuous version of this model was introduced by Zhang to study
the propagation of activated energies \cite{ZHANG_1}.
In contrast to the BTW model the Zhang model is a non Abelian model, 
i.e., the steady state configurations depend on the sequence in 
which unstable sites are toppled (see \cite{PRIEZ_2,IVASH_2} and
references therein).
Despite the different microscopic dynamics both models
are expected to belong to the same universality class
(see for instance \cite{DIAZ_2}).
Up to now nobody has proved this assumption by
direct measurements of the avalanche exponents on large
lattice sizes which reduces the finite size effects sufficiently.
We consider the Zhang model on lattice sizes which are significantly
larger than those sizes used in previous investigations 
\cite{ZHANG_1,JANOSI_1,PIETRO_1,DIAZ_1}.
The energy distribution  $p(E)$ which characterizes
the static properties of the model
is concentrated around 
$z$ distinct peaks where $z$ is the number of nearest neighbors.
We show that the peaks are located 
at multiples of $\frac{z+1}{z^2}$ 
and that the height of the peaks grow with increasing system size.
Numerical simulations of the two dimensional square and honeycomb
lattices confirm this result.
We also investigated the avalanche distributions on
lattice sizes up to  $L=2048$.
A finite size analysis of the exponents of the 
avalanche distributions yields values 
which corresponds to that of the BTW model.

\section{Model}

We consider a two dimensional square lattice of linear size $L$.
A continuous value $E_{i,j}\ge0$ representing the energy 
is associated to each lattice site $(i,j)$.
The boundary sites are fixed to zero ($E(boundary)=0$) for all times.
A configuration $\{E_{i,j}\}$ is stable if $E_{i,j}<E_c$
for all lattice sites $(i,j)$.
For the sake of simplicity we choose in all simulations $E_c=1$.
A quantum of energy $\delta$ is added to a
randomly chosen lattice site $(i,j)$, i.e.,
\begin{equation}
E_{i,j}\;\to\;E_{i,j}\,+\,\delta.
\label{eq:perturbation}
\end{equation}
In the case that due to this perturbation a site exceeds the
critical value $E_c$ an activation event will occur and the
critical site relaxes to zero and the energy is added
to the next neighbors, i.e.,
\begin{equation}
E_{i,j}\;\to\;0,
\label{eq:relaxation_1}
\end{equation}
\begin{equation}
E_{i,j,NN}\;\to\;E_{i,j,NN}\;+\;\frac{E_{i,j}}{z},
\label{eq:relaxation_2}
\end{equation}
where $z$ denotes the number of next neighbors.
In that way the transferred energy may activate the neighboring
sites and thus an avalanche of relaxation events may take place.
Energy may leave the system only at the boundary.

In our simulations we use various values of the input 
energies out of the interval $\delta \in ]0,E_c]$.
In the case of $\delta \to 0$ all lattice sites grow parallel.
In order to implement this different perturbation
process one has to find the site with the largest
energy $E_{max}$ and then increment all sites by $E_c-E_{max}$.
In this case the Zhang model is identical
with the conservative limit of the ``spring block'' 
model of Christensen and Olami \cite{CHRIS_1}.

\begin{figure}
 \begin{minipage}[t]{8.6cm}
 \epsfxsize=8.6cm
 \epsfysize=8.6cm
 \epsffile{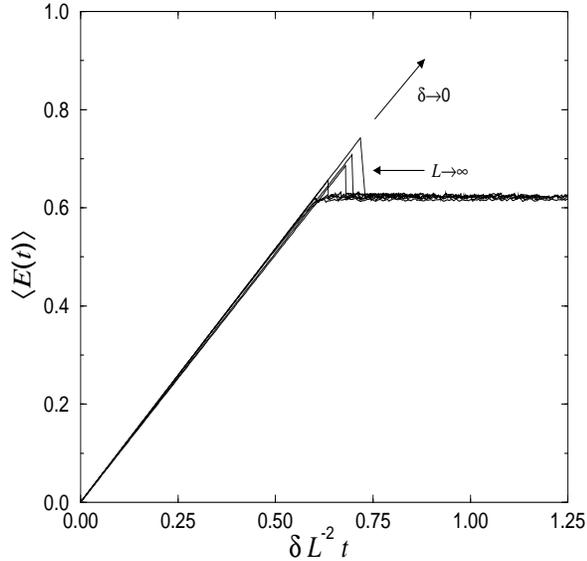} 
 \caption{The average energy $\langle E (t)\rangle$ as a function
of the rescaled time $\tau=\delta L^{-2} t$
for $L\le 512$ and various values of $\delta$.
 \label{steady_state}} 
 \end{minipage}
\end{figure}

The concept of self-organized critical systems
refers to driven systems which organize themselves 
into a steady state.
We consider the average energy 
\begin{equation}
\langle E(t) \rangle \; = \; \frac{1}{L^2} \, \sum_{i,j} \, E_{i,j}(t),
\label{eq:order}
\end{equation}
to check if the system has reached the steady state.
Starting with an empty lattice 
we consider the growth of the pile.
In the beginning all sites are subcritical, i.e., $E_{i,j} \ll E_c$
and no toppling event occurs.
Here, no relaxation process takes place (non avalanche regime) 
and the average energy
increases linear in time (see Fig.~\ref{steady_state}).
With further perturbations the average energy is still growing
until one site reaches the critical value $E_c$.
Now the behavior of the system changes 
and toppling processes occur  (avalanche regime).
After a certain time the average energy reaches a constant
value $\langle E \rangle$ which characterizes the steady state.

In Fig.~\ref{steady_state} the average energy is plotted 
as a function of the rescaled time $\tau=\delta \, L^{-2} \, t$.
One can see a data collapse of all curves corresponding to different 
values of $L$ and $\delta$.
Deviations from the collapse occur only at the point $\tau \approx 0.63$
where the behavior changes from the non avalanche regime to the 
avalanche regime, characterized by a constant average energy.
The avalanche regime occurs when the fluctuations of the energies 
are greater than the difference
of the critical energy $E_c$ from the average energy $\langle E (t)\rangle$,
i.e., when
\begin{equation}
\FL
\sqrt{ \langle E^2(t) \rangle \;-\, \langle E (t) \rangle ^2 } \; \geq  \;
E_c \, - \,  \langle E(t) \rangle. 
\end{equation}

Decreasing $\delta$ reduces the fluctuations and the critical time 
tends to $\tau_c=1$.
Larger system sizes result in a decreasing critical time.

We consider the system for lattice sizes 
$L\in \{64, 128, 256, 512, 1024, 2048\}$ (in the case of $\delta=0$
the maximum lattice size is $L=1024$). 
Starting with an empty lattice the system will be equilibrated 
after $L^2 \delta^{-1}$ perturbations.
In order to provide a sufficient statistics
all measurements are averaged over at least $10^6$ non zero avalanches.

\section{Energy Distribution}

\begin{figure}
 \begin{minipage}[t]{8.6cm}
 \epsfxsize=8.6cm
 \epsfysize=8.6cm
 \epsffile{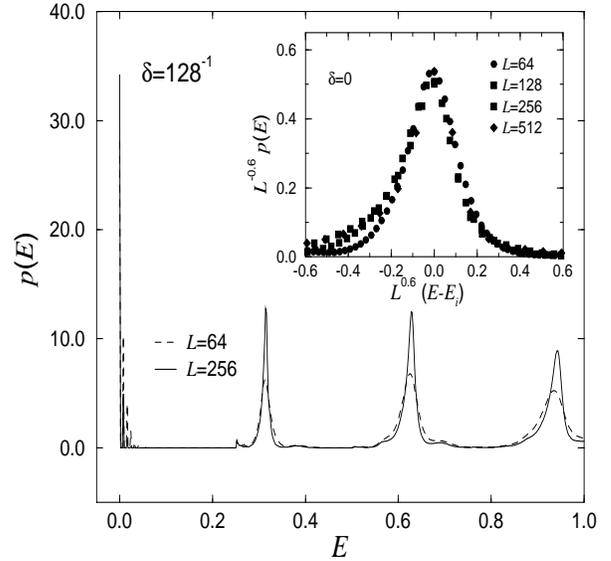} 
\caption{The probability distribution $p(E)$ for different system sizes.
The inset displays the scaling plot of the third maximum of $p(E)$.
 \label{energy_dist}}
 \end{minipage}
\end{figure}

We measured the energy distribution $p(E)$ 
in the steady state for  
$\delta \in \{0,128^{-1},8^{-1}, 1\}$
and various system sizes $L$.
In Fig.~\ref{energy_dist} the distribution $p(E)$ is plotted
for different system sizes.
The distribution is concentrated around four distinct peaks.
It was assumed in previous works \cite{ZHANG_1,PIETRO_1}
that the finite spreads of the peaks are caused by ``intrinsic dynamical
fluctuations''.
As one can see from Fig.~\ref{energy_dist} 
the peaks grow and the spreads of the peaks decrease  
with increasing system size $L$.

We found that the maximum $p_{max}(E)$ of each peak scales with
the system size as
\begin{equation}
p_{max}(E) \, \sim \,L^y,
\label{eq:E_MAX_SCALING}
\end{equation}
with $y\approx0.6$.
Since the distribution $p(E)$ is normalized 
we assume that the peaks scale
in the horizontal direction as $L^{-y}$.
The location of the maxima of the distribution $p(E)$ depend
slightly on the system size $L$.
In order to produce a scaling plot this drift has to be taken into
account.
In the inset of Fig.~\ref{energy_dist} we plot $L^{-y} p(E)$ 
as a function of $L^y (E-E_{max}(L))$ and get a satisfying data collapse.
The peaks of the energy distribution $p(E)$ grow to infinity 
and the spread of each peak vanishes for $L\to\infty$. 
In the case of an infinite
system the energy distribution $p(E)$ in the steady state
is given by
\begin{equation}
p(E)\,=\,\sum_{i=0}^3 \,f_i \;\delta(E-E_i),
\label{eq:p_infty}
\end{equation} 
where $f_i$ denotes the statistical weight and $E_i$ denotes 
the location of the $\delta$-peaks.

One can calculate
the discrete values of the energies $E_i$ in the following way 
\cite{HUCHT}:
Suppose that the energies are already discretized with the allowed values
\begin{equation}
E \in \{ 0 ,\, E_0, \,2 \,E_0, \, 3 \,E_0,~...~, \,n\, E_0,~...\}.
\label{eq:E_values}
\end{equation}
Then a maximum value of $n$ exists with
\begin{equation}
n_{max}\,E_0\, \;\le \;E_c \;< \; (n_{max}+1)\, E_0.
\end{equation}
The critical energy $E=(n_{max}+1) E_0$ relaxes and $\frac{E}{z}$  
should be equal to $E_0$, i.e.,
\begin{equation}
\frac{(n_{max}+1) \,E_0}{z} \; = \;E_0.
\label{eq:det_n_max}
\end{equation}
In this way the number of peaks equals
the lattice coordination number $n_{max}+1=z$.
Based on his numerical investigations of different lattice types 
D\'{\i}az-Guilera has already proposed this relation \cite{DIAZ_1}.

\begin{figure}
 \begin{minipage}[t]{8.6cm}
 \epsfxsize=8.6cm
 \epsfysize=8.6cm
 \epsffile{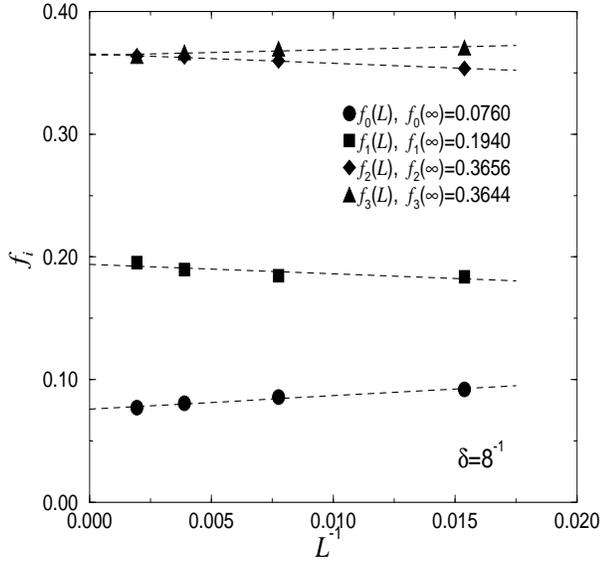} 
 \caption{The statistical weights $f_i$ as a function of the inverse 
	  system size $L^{-1}$ for $\delta=8^{-1}$. The 
	  values $f_i(\infty)$ are obtained
	  by an extrapolation to the vertical axis.
 \label{stat_weight}}  
 \end{minipage}
\end{figure}

\begin{table}[b]
\caption{Statistical weights of the energy distribution.}
\label{TABLE_F}
\begin{tabular}{cccccc}
  & $\delta=0$ & $\delta=128^{-1}$  & $\delta=8^{-1}$  & $\delta=1$ & BTW \cite{PRIEZ_1} \\  
\tableline
$f_0$  & 0.077  & 0.076 & 0.076 & 0.077 & 0.074 \\
$f_1$  & 0.197  & 0.196 & 0.194 & 0.195 & 0.174 \\
$f_2$  & 0.365  & 0.362 & 0.366 & 0.364 & 0.306 \\ 
$f_3$  & 0.362  & 0.366 & 0.364 & 0.364 & 0.446 \\
\end{tabular}
\end{table}

Starting with a stable configuration
one perturbs the system until one site becomes critical, 
i.e., one adds $\Delta E=E_c-n_{max}\,E_0$ on each lattice site 
(this is correct for $\delta \to 0$).
The energy of a given site is now $E=n \, E_0+\Delta E$.
The critical site relaxes and $\frac{E_c}{z}$ is added to
the $z$ next neighbors of this site.
Arguing that the new energy is the next allowed energy value 
$E=(n+1) E_0$ one gets the relation
\begin{equation}
E_0 \; = \; \frac{E_c}{n_{max}+1} \, \frac{z+1}{z} 
\;=\; E_c\,\frac{z+1}{z^2}.
\label{eq:E_0}
\end{equation}
Note that the discretization of the energies
is independent of the dimension of the system. 
The relevant term is the lattice coordination number $z$.
This is in contrast to the conclusions drawn from previous investigations.
These based only on simulations of square lattices in different 
dimensions $d$ where the 
coordination number is given by $z=2d$ \cite{JANOSI_1}.

We compare Eq.~(\ref{eq:E_0}) with the results obtained from
computer simulations. 
In $d=2$ we found that $E_0 \approx 0.3149$ for $\delta=128^{-1}$,
 $E_0 \approx 0.3153$ for $\delta=0$,
 $E_0 \approx 0.3145$ for $\delta=8^{-1}$, and 
 $E_0 \approx 0.3140$ for $\delta=1$ 
which are in good agreement with Eq.~(\ref{eq:E_0}).
We also measured the energy distribution of a honeycomb lattice in two
dimensions ($z=3$) and found for $\delta=128^{-1}$
the value $E_0 \approx 0.443$ which corresponds very good to the
exact value $E_0=0. \bar 4$.
Pietronero {\it et al}.~have investigated the
$d=3$ Zhang model on a square lattice and found 6 peaks 
in the energy distribution \cite{PIETRO_1}. 
We measured the average distance between two peaks from
the Fig.~2 of \cite{PIETRO_1} and determined
in this way $E_0=0.190$ which agrees with $E_0=0.19 \bar 4$ obtained 
from Eq.~(\ref{eq:E_0}).

Furthermore, we determined the statistical weights $f_i$ of the
energy distribution [Eq.~(\ref{eq:p_infty})].
We divided the interval $[0,E_c]$ in four 
parts and measured in each part the area $f_i(L)$ under the curve $p(E)$
for various system sizes $L$.
The statistical weights $f_i$ are given by an extrapolation to 
$L\to \infty$ (see Fig.~\ref{stat_weight}) and the
obtained values are listed in Table~\ref{TABLE_F}.
Analogous to the locations of the peaks the statistical
weights do not depend on the input energy $\delta$.
On the other hand one can see that the values differ
from those of the BTW model which are known exactly \cite{PRIEZ_1}.

\begin{figure}
 \begin{minipage}[t]{8.6cm}
 \epsfxsize=8.6cm
 \epsfysize=8.6cm
 \epsffile{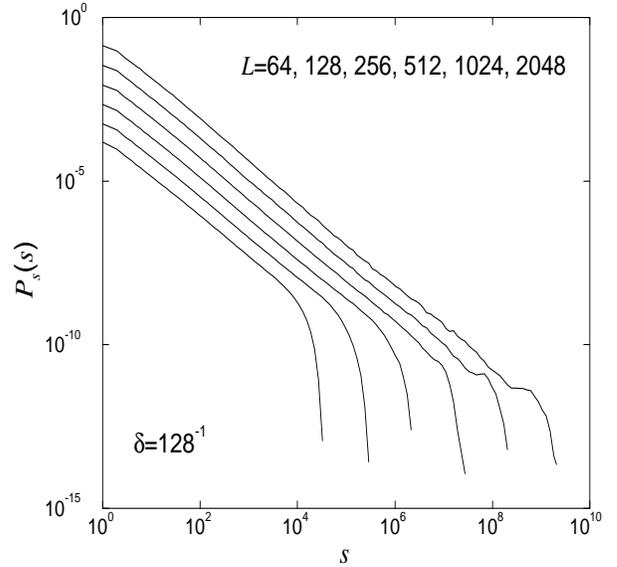} 
 \caption{The probability distribution $P_s(s)$ for different
	  system sizes for $\delta=128^{-1}$. The curves 
          for $L<2048$ are shifted in the downward direction.
 \label{size_prob}}
 \end{minipage}
\end{figure}

Pietronero {\it et al}.~\cite{PIETRO_2} introduced a renormalization 
group approach for sandpile models where
the density of the critical sites determines 
the fixed point of the renormalization transformation.
Here, the density of the critical sites corresponds to
the statistical weight $f_3$ in the sense that any perturbation
of a coarse grained particle ($E_0$) leads to a relaxation
event.
Following our results both models are characterized by different 
fixed points and thus one might expect that both models belong to
different universality classes.
But one has to emphasize that this renormalization group approach
and its improvement by Ivashkevich \cite{IVASH_2} neglects
fluctuations at the steady state.
Due to this ``mean-field-type approximation'' \cite{KATORI} we think that
the different critical densities of the Zhang and the BTW model
cannot lead to an answer of the universality question.

\begin{figure}
 \begin{minipage}[t]{8.6cm}
 \epsfxsize=8.6cm
 \epsfysize=8.6cm
 \epsffile{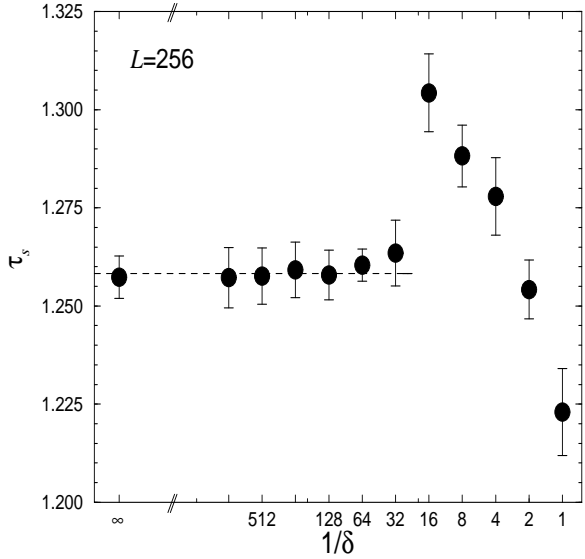} 
 \caption{The values of the exponent $\tau_s$ as a function
	  of the input energy $\delta$ for a fixed system size $L$.
	  Note that the values of the exponent are independent of $\delta$ in the
	  limit $\delta \ll E_c$. 
 \label{tau_s_delta}}
 \end{minipage}
\end{figure}

\section{Avalanche Distributions}

In this section we examine the probability distribution
of an avalanche of size $s$, area $s_d$, duration $t$,
and radius $r$,
where $s$ denotes the total number of toppled sites,
$s_d$ is the number of distinct sites which corresponds
to the area of an avalanche.
The duration $t$ of an avalanche is equal to
the number of update sweeps needed until all sites are
stable again.
The linear size of an avalanche $r$ is measured via the
radius of gyration of the avalanche cluster.
In the critical steady state the corresponding probability 
distributions should obey power-law behavior 
characterized by exponents $\tau_s$, $\tau_d$,  $\tau_t$, and $\tau_r$
according to
\begin{equation}
P_s(s) \, \sim \, s^{-{\tau_s}},
\label{eq:prob_size}
\end{equation}
\begin{equation}
P_d(s_d) \, \sim \, {s_d}^{-{\tau_d}},
\label{eq:prob_distinct}
\end{equation}
\begin{equation}
P_t(t) \, \sim \, t^{-{\tau_t}},
\label{eq:prob_duration}
\end{equation}
\begin{equation}
P_r(r) \, \sim \, r^{-{\tau_r}}.
\label{eq:prob_radius}
\end{equation}

The distribution $P_s(s)$ is plotted in Fig.~\ref{size_prob} 
for $\delta=128^{-1}$ and various system sizes $L$.
All curves fit in the middle region to a straight line
and the corresponding exponents are determined via 
regression of this region.
First we investigate whether the exponents depend on the
input energy $\delta$ and second we examine 
how the finite system size affects the results.
Figure \ref{tau_s_delta} shows the 
exponent $\tau_s$ for $L=256$ and for various values
of $\delta$.
In the limit $\delta \ll E_c=1$ the exponents
are independent of the input energy.
This behavior changes abrupt for $\delta \ge 32^{-1}$ where
the exponent displays a complex $\delta$ dependence.
In the following we focus our attention on the limit
$\delta \ll E_c$.  

\begin{figure}
 \begin{minipage}[t]{8.6cm}
 \epsfxsize=8.6cm
 \epsfysize=8.6cm
 \epsffile{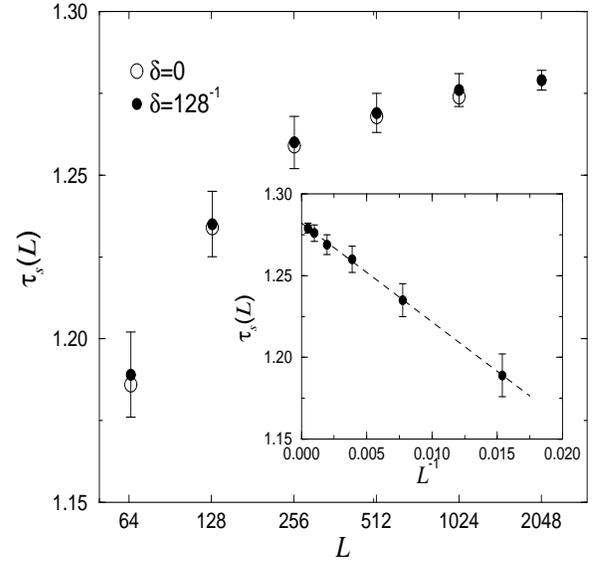} 
 \caption{The system size dependence of the exponent $\tau_s$ for 
	$\delta=0$ and $\delta=128^{-1}$.
        The inset displays the determination of $\tau_{\infty}$ according
        to Eq.~(\protect\ref{eq:tau_L}) (dashed line).
 \label{tau_s}}
 \end{minipage}
\end{figure}

The exponents $\tau_s$ corresponding to different values 
$L$ and $\delta$ are plotted in Fig.~\ref{tau_s}.
Significant differences between the values of the
exponents $\tau_s(L,\delta=0)$ and $\tau_s(L, \delta=128^{-1})$ 
are caused by the system size only and not by the input energy.
Both exponents tend to $\tau_s \approx 1.28$ with increasing $L$.
In order to determine the exact value of the exponent $\tau_s$
we assume that its system size dependence is given by
\begin{equation}
\tau_{s}(L) \;=\; \tau_{s} \,+\,\frac{\mbox{const}}{ L^{x}}.
\label{eq:tau_L}
\end{equation}
We tried several values of $x$ and got the best results for $x=1$,
i.e., the finite size effects are of the relative magnitude
of the boundary ($\sim L^{-1}$).
In the inset of Fig.~\ref{tau_s}  the exponents $\tau_{s}(L)$ are 
plotted as a function of the inverse system size.
The exponent $\tau_{s}$ is given by an extrapolation
to $L\to\infty$ which yields $\tau_{s}=1.282 \pm 0.01$.

\begin{figure}
 \begin{minipage}[t]{8.6cm}
 \epsfxsize=8.6cm
 \epsfysize=8.6cm
 \epsffile{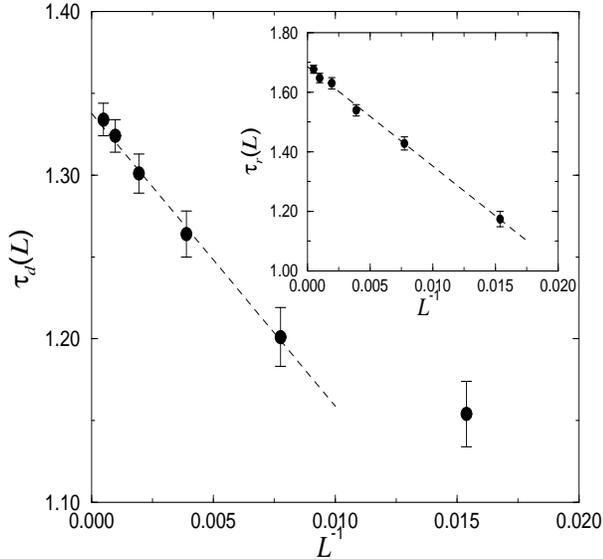} 
 \caption{The values of the exponent $\tau_d$ and $\tau_r$ as a function
	  of the inverse system size $L^{-1}$ for $\delta=128^{-1}$. 
	  The dashed lines correspond to the extrapolation according 
	  to Eq.~(\protect\ref{eq:tau_L})
 \label{tau_d_r}}
 \end{minipage}
\end{figure}

The exponents of the avalanche probability distribution of the area
and radius are characterized by the same finite-size corrections.
The exponents corresponding to different system sizes
are plotted in Fig.~\ref{tau_d_r}.
Except of the deviation for $L=64$ in the case of the
exponent $\tau_d$ both exponents depend on the inverse system 
size (corresponding to Eq.~(\ref{eq:tau_L})).
From the extrapolation to the infinite system size we obtain
the values $\tau_d=1.338 \pm 0.015$ and 
$\tau_r = 1.682 \pm 0.018$, respectively.
The finite size dependence explains why 
lower values of the exponents were reported in
previous works based on numerical simulations 
of one system size only (see for instance \cite{JANOSI_1}).

\begin{figure}
 \begin{minipage}[t]{8.6cm}
 \epsfxsize=8.6cm
 \epsfysize=8.6cm
 \epsffile{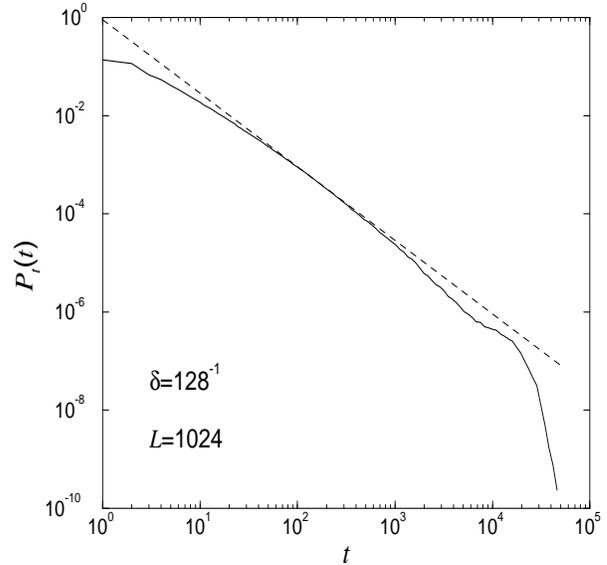} 
 \caption{The probability distribution $P_t(t)$ for a fixed
	  system size. The dashed line corresponds to a power-law
	  with the exponent of the BTW model $\tau_t=\frac{3}{2}$ 
	  (see \protect\cite{LUEB_2}).
 \label{duration_prob}}
 \end{minipage}
\end{figure}

The finite size analysis described above fails in the case of the duration
exponent $\tau_t$.
Here, the probability distribution exhibits a finite curvature
which makes it impossible to determine the exponent via 
regression (see Fig.~\ref{duration_prob}).
Using a momentum-space analysis of the corresponding
Langevin equations D\'{\i}az-Guilera showed that
the dynamical exponent of the BTW and Zhang's model 
is given by $z=(d+2)/3$ \cite{DIAZ_2}.
This result allows to determine the exponent $\tau_t$ 
because the exponents $z, \tau_t$, and $\tau_r$ have
to fulfill the scaling relation 
(see for instance \cite{LUEB_2}) 
\begin{equation}
{z}\;=\;\frac{\tau_r-1}{\tau_t-1}.
\label{eq:gam_tr}
\end{equation}
Using the above value of $\tau_r$ and $z=\frac{4}{3}$
for the two-dimensional model we obtain the 
value $\tau_t=1.512\pm0.014$.

Recently, it has been shown numerically that the exponents of the 
BTW model are consistent with the values $\tau_t=\frac{3}{2}$,  
$\tau_d=\frac{4}{3}$, and  $\tau_r=\frac{5}{3}$ \cite{LUEB_2}.
Because of the lack of a scaling relation the exact value of $\tau_s$
is still unknown but the authors estimate the 
value $\tau_s=1.293\pm0.009$.
These values are in agreement with our results, strongly 
suggesting  that both models are characterized by 
the same exponents.

Note that we determined the exponents of the Zhang model 
for the limit $\delta \ll E_c$ only.
The measurements for a fixed system size and larger values
of the input energy $\delta$ yield different values of the
exponents (see Fig.~\ref{tau_s_delta}).
But this does not mean that the exponents of the infinite
system size depend on $\delta$.
It is also possible that the finite size behavior [Eq.~(\ref{eq:tau_L})] 
changes outside the limit $\delta \ll E_c$.
Further work has to be done to examine 
how the finite system size affects the values of the avalanche 
exponents for $\delta \approx E_c$.

\section{Conclusions}

We have studied numerically the static and dynamical properties
of the non Abelian Zhang model on large system sizes.
The steady state energy distribution is concentrated 
around $z$ distinct peaks which are located at multiples 
of $\frac{z+1}{z^2}$, where $z$ denotes the lattice 
coordination number.
The statistical weights of the peaks are independent of the
input energy $\delta$ but differ from those of the BTW model.
A finite size analysis of the avalanche 
probability distributions in the limit $\delta \ll E_c$ yields 
exponents which are in agreement with the values of the 
exponents of the BTW model.
Both models belongs to the same universality class, i.e., both 
models displays the same large scale behavior,
characterized by the avalanche exponents.

\acknowledgments
I would like to thank S.~S.~Manna for having interested me in
the Zhang model and for helpful discussions
and K.~D.~Usadel for a critical reading of 
the manuscript.
This work was supported by the
Deutsche Forschungsgemeinschaft through
Sonderforschungsbereich 166, Germany.

\end{document}